\documentclass{ws-ijmpd}
\usepackage{latexsym}
\usepackage{amssymb}
\usepackage{amsfonts}
\usepackage{amsmath}
\usepackage{graphicx}

\begin{document}

\markboth{M.L. Ruggiero, A. Tartaglia, L. Iorio} {Doppler Effects
from Bending of Light Rays in Curved Space-Times}

%
\catchline{}{}{}{}{}
%

\title{DOPPLER EFFECTS FROM BENDING OF LIGHT RAYS IN CURVED SPACE-TIMES}

\author{MATTEO LUCA RUGGIERO}

\address{Dipartimento di Fisica, Politecnico di Torino, \\  Corso Duca degli Abruzzi 23, 10129 Torino, Italy \\
INFN, Sezione di Torino \\ Via Pietro Giuria 1, 10125 Torino,
Italy \\ matteo.ruggiero@polito.it}

\author{ANGELO TARTAGLIA}

\address{Dipartimento di Fisica, Politecnico di Torino, \\  Corso Duca degli Abruzzi 23, 10129 Torino, Italy \\
INFN, Sezione di Torino \\ Via Pietro Giuria 1, 10125 Torino,
Italy \\ angelo.tartaglia@polito.it}

\author{LORENZO IORIO}

\address{Viale Unit\`{a} d'Italia 68, 70125 Bari, Italy  \\ lorenzo.iorio@libero.it}

\maketitle
\begin{history}
\received{Day Month Year} \revised{Day Month Year} \comby{Managing
Editor}
\end{history}

\begin{abstract}
We study Doppler effects in curved space-time, i.e. the frequency
shifts induced on electromagnetic signals propagating in the
gravitational field. In particular, we focus on the frequency
shift due to the bending of light rays in weak gravitational
fields. We consider, using the PPN formalism, the gravitational
field of an axially symmetric distribution of mass. The zeroth
order, i.e. the sphere, is studied then passing to the
contribution of the quadrupole moment, and finally to the case of
a rotating source. We give numerical estimates for situations of
physical interest, and by a very preliminary analysis, we argue
that analyzing the Doppler effect could lead, in principle, in the
foreseeable future, to the measurement of the quadrupole moment of
the giant planets of the Solar System.
\end{abstract}



\section{Introduction}

\label{sec:intro}

Einstein's theory of gravity, General Relativity (GR), has passed
all observational tests with excellent results, at least at the
scale of the Solar System and in many binary compact systems;
however as it is well known, some problems arise with the recent
cosmological observations. In fact, the evidence of the
acceleration of the Universe is supported by experimental data
deriving from different tests: i.e., from Type-Ia Supernovae, from
CMWB and from the large scale structure of the universe
\cite{c1,c2,c3,c4,c5}. However, GR is not able to provide a
theoretical explanation of these experimental facts unless some
\textit{exotic and invisible energy} is admitted to exist in the universe (%
\textit{Dark Energy}). Also for these reasons, the reliability of
the theories of gravity alternative to GR is of great interest
today. Of course, all these theories must agree with the known
tests of gravity performed in the weak field and slow motion
limit, in the Solar System: the standard tool used for formulating
(metric) gravity theories in this context is the Parameterized
Post-Newtonian (PPN) formalism \cite{will}. Different theories are
characterized by different values of the coefficients appearing in
front of the post-Newtonian metric potentials, and the formalism
determines the values of these parameters for each theory under
consideration. Consequently, the formalism allows for a direct
comparison of the competing theories with each other, and with the
results of the experiments and observations. In particular, Solar
System tests limit the range of the values that the PPN parameters
may assume, thus excluding some theories of gravity and
quantitatively limiting the deviations from GR \cite{will05}.

The study of the Doppler effects in the space-time around a
gravitating body was carried out by many authors in the past (see,
for instance,
\cite{gron,bertotti92,krisher93,kopeikin99,kopeikin02} and
references therein), both for pedagogical purposes and for
studying the feasibility of actual experiments aimed at detecting
the effects themselves. In particular, in \cite{kopeikin99},
Kopeikin and Sch\"{a}fer carefully studied  the covariant theory
of propagation of light in very general gravitational fields, and
the various phenomena related, among which, the frequency shift.
Indeed, as it is clearly explained in \cite{iess99}, the Doppler
frequency shift, the deflection of light and the time delay of
electromagnetic signals are related to each other, in the sense
that they might be thought of as "different manifestation of the
same aspect of the [...] gravitational field". However, their
measurements require different experimental techniques and,
moreover, the interpretation of the measured quantities is quite
different. In particular, the Doppler frequency shift is a
relativistic invariant, which, roughly speaking, corresponds to
the projection of the photons' four-momentum along the world-line
of the observer, and, consequently, it can be defined without
reference to any particular coordinate frame.

As for the measurement technique, in actual Doppler experiments a
radio beam is transmitted from the Earth to the receiver located
aboard a spacecraft; the radio beam is coherently transponded and
then sent back to the Earth, where the received frequency is
measured with great accuracy, usually by means of hydrogen masers.
The comparison of the transmitted and received frequencies gives
the measurement of the Doppler shift.

In the past, there have been different proposals and different
experimental tests of the relativistic frequency shift (see
\cite{krisher93} and references therein), and, indeed, very
recently, a test of GR was performed by studying the Doppler
effect from the Cassini spacecraft. In particular, a measurement
of the post-Newtonian parameter $\gamma$, which is equal to $1$ in
GR, was carried out: as a result of the experiment, $\gamma=1 +
(2.1 \pm 2.3) \times 10^{-5}$, thus sensibly limiting the
deviations from GR \cite{bertotti03}.

Here we apply the study of the Doppler frequency shift to some
situations that may be of interest, both at the Solar System and
astrophysical scale. In particular, we focus on Doppler effects
originating from the bending of light rays in curved space-times
(which will be measured, with great accuracy by the LATOR mission
\cite{lator}), in order to see if these effects could be used to
constrain the PPN parameters and, on the other hand, we argue that
they can be used as probes to determine some astrophysical
properties of interest of the celestial bodies. Numerical
estimates of the magnitude of these effects are given, and they
are compared to the current accuracies; furthermore, the
characteristics of the effects are examined, in order to suggest
possible detection techniques.

The paper is organized as follows: in Section \ref{sec:problem} we
briefly outline the general framework used to study the Doppler
frequency shift, according to previous works; in Section
\ref{sec:application} we apply the
general approach to some metrics of physical interest; in Section \ref%
{sec:estimates} we consider some prototypical situations and give
numerical estimates for them; finally, the possibility of
measuring the effects is discussed in Section
\ref{sec:conclusions}.


\section{The Problem}

\label{sec:problem}

In this section we outline the theoretical framework which allows
to study the frequency shifts of electromagnetic signals
propagating in a gravitational field. In doing so, we follow the
approach of Bertotti and Giampieri \cite{bertotti92}, which is in
agreement for the cases under consideration, with the very general
approach developed by Kopeikin and Sch\"{a}fer \cite{kopeikin99}.

Let us consider an emitter of monochromatic electromagnetic waves,
which emits with proper frequency $\nu _{E}$. The emitter is
moving in the gravitational field of a massive source; let us call
$\bm{u}_{E}$ its four-velocity, evaluated at the (coordinate) time
of emission of a signal when the emitter is at the position
$\vec{\bm{x}}_{E}$ (the massive source is located at the origin of
our reference frame). The problem is to evaluate the frequency
$\nu _{R}$ measured by a receiver at $\vec{\bm{x}}_{R}$, whose
four-velocity is $\bm{u}_{R}$ \footnote{%
Greek indices run from 0 to 3, Latin indices run from 1 to 3; the
space-time metric has signature $(-1,1,1,1)$, and we use units
such that G=c=1; nonetheless, for the sake of clarity, we
re-introduce physical units in the final formulae; boldface
arrowed letters, refer to vectors in the three-dimensional space;
boldface letters refer to four-vectors.}.

We may write the four-velocities in terms of the world-lines of the emitter (%
$\bm{x}_{E}$) and receiver ($\bm{x}_{R}$)
\begin{equation}
\bm{u}_{E}=\frac{d\bm{x}_{E}}{ds_E}\ \ \ \
\bm{u}_{R}=\frac{d\bm{x}_{R}}{ds_R}. \label{eq:fourvel1}
\end{equation}%
Of course, in terms of the parameters $s_E,s_R$ along the
world-lines:
\begin{equation}
\bm{x}_{E}(s_E)\equiv \left( x_{E}^{0}(s_E),\vec{\bm{x}}_{E}(s_E)\right) ,\ \ \ %
\bm{x}_{R}(s_R)\equiv \left(
x_{R}^{0}(s_R),\vec{\bm{x}}_{R}(s_R)\right) . \label{eq:fourvel2}
\end{equation}%
Furthermore let $\bm{\xi}(\lambda )=\bm{\xi}_{0}(\lambda )+\delta \bm{\xi}%
(\lambda )$ be the world-line of the emitted photon, where
$\bm{\xi}_{0}$ is the trajectory that the photon would have if it
were emitted in flat space-time (i.e. a straight line) and $\delta
\bm{\xi}$ is the perturbation induced by the gravitational field.
This world-line is parameterized in such a way that the photon is
emitted at $\vec{\bm{x}}_{E}(s_{E})$ when $\lambda =-l_{E}$, and
it is received at $\vec{\bm{x}}_{R}(s_{R})$ when $\lambda =l_{R}$.

Let us define the four-vector tangent to the photons' world-line
\begin{equation}
\bm{p}=\frac{d\bm{\xi}(\lambda )}{d\lambda }=\frac{d\bm{\xi_0}(\lambda )}{%
d\lambda }+\frac{d\delta \bm{\xi}(\lambda )}{d\lambda }\doteq \bm{p}%
_{0}+\delta \bm{p}.  \label{eq:formom1}
\end{equation}%
where $\bm{p}_0$ is the flat space-time term and $\delta \bm{p}$
is the perturbation induced by the gravitational field. If we set
$l=l_{E}+l_{R}$ the flat space-time term $\bm{p}_{0}$ can be
written in the form
\begin{equation}
\bm{p}_{0}=\left(
1,\frac{\vec{\bm{x}}_{R}-\vec{\bm{x}}_{E}}{l}\right) \doteq \left(
1,\hat{\bm{n}}\right),  \label{eq:defpzero1}
\end{equation}%
where $\hat{\bm{n}}$ is a unit vector along the spatial path of
the photons in flat space-time.

The frequencies in the receiver and emitter reference frames are
related by the following formula \cite{straumann,sch}
\begin{equation}
\frac{\nu _{R}}{\nu _{E}}=\frac{\bm{u}_{R}\cdot \bm{k}_{R}}{\bm{u}_{E}\cdot %
\bm{k}_{E}},  \label{eq:freq1}
\end{equation}
where $\bm{k}$ is the photons' wave vector. Indeed, in general,
the frequency $\nu_p$ of a photon measured by an observer at a
point $P$ is found by projecting $\bm{k}$ onto the observer's
four-velocity:
\begin{equation}
2\pi \nu_p = \bm{k}_P \cdot \bm{u}_P. \label{eq:proj11}
\end{equation}
Now, we can expand the components of the wave vector about their
flat space-time values \cite{krisher93}:
\begin{equation}
\bm{k}= \omega \left(\bm{p}_0+\delta\bm{p} \right) = \left[-\omega
\left(1+\delta p^0\right), \omega\left(\hat{\bm{n}}+\delta
\vec{\bm{p}} \right) \right]. \label{eq:wavec11}
\end{equation}
Hence, eq. (\ref{eq:freq1}) can be written as
\begin{equation}
\frac{\nu _{R}}{\nu _{E}}=\left( \frac{\bm{u}_{R}\cdot \bm{p}_{0}}{\bm{u}%
_{E}\cdot \bm{p}_{0}}\right) \left( \frac{1+\frac{\bm{u}_{R}\cdot \delta %
\bm{p}_{R}}{\bm{u}_{R}\cdot \bm{p}_{0}}}{1+\frac{\bm{u}_{E}\cdot \delta %
\bm{p}_{E}}{\bm{u}_{E}\cdot \bm{p}_{0}}}\right) , \label{eq:freq2}
\end{equation}%
where the fact that $\bm{p}_0 |_E=\bm{p}_0 |_R$ has been
exploited. From (\ref{eq:freq2}) we then obtain
\begin{equation}
\ln \left( \frac{\nu _{R}}{\nu _{E}}\right) =\ln \left( \frac{\bm{u}%
_{R}\cdot \bm{p}_{0}}{\bm{u}_{E}\cdot \bm{p}_{0}}\right) +\ln
\left(
\frac{1+\frac{\bm{u}_{R}\cdot \delta \bm{p}_{R}}{\bm{u}_{R}\cdot \bm{p}_{0}%
}}{1+\frac{\bm{u}_{E}\cdot \delta \bm{p}_{E}}{\bm{u}_{E}\cdot \bm{p}_{0}}}%
\right) .  \label{eq:freq3}
\end{equation}%

We see that the frequency shift (\ref{eq:freq3}) is made of two
contributions. The first one depends on the velocities of the
emitter and receiver and, also, on the gravitational field at the
emission and reception points. In particular, it is possible to
show \cite{krisher93} that the ordinary Doppler effect
(transversal and longitudinal) and the gravitational frequency
shift come from this term.

The second term depends, again, on the velocities and on the
gravitational field, but, furthermore, it depends also on the
gravitational perturbations of the world-lines of the photons. In
the following, we focus on this term only. Also, we assume that
space-time is flat at infinity, which amounts to say that, in the
limit for $l_{E},l_{R}\rightarrow \infty $, we may neglect the
gravitational field at the emission and reception point.
Consequently the first contribution accounts for the special
relativistic  effects (and,
to lowest order, it is $O(v)$\footnote{%
Since in our units $c=1$, $O(v^{n})$ means $O(v^{n}/c^{n})$, in
other words we use approximations with respect to the small
parameter $v/c$, where $v$ is the scalar velocity.}, where $v$ is
here the relative velocity of the emitter and the receiver), while
the second term takes into account the gravitational effects along
the world line.

On neglecting the gravitational field both at the emission and
reception points (indices are raised and lowered by means of the
Minkowski tensor $\eta _{\mu \nu }$), we may write the
four-velocities of the emitter and the receiver
\begin{equation}
\bm{u}_{E/R}=\gamma _{E/R}\left( 1,\vec{\bm{v}}_{E/R}\right) ,
\label{eq:veloc1}
\end{equation}%
where
\begin{equation}
\gamma _{E/R}=\frac{1}{\sqrt{1-v_{E/R}^{2}}}.  \label{eq:veloc2}
\end{equation}%

So, the  contribution due to the perturbation of the photons'
paths ("gravitational" contribution) reads
\begin{equation}
\ln \left( \frac{\nu _{R}}{\nu _{E}}\right) _{gr}=\left[ \ln \left( 1+\frac{%
\delta p^{0}-\vec{\bm{v}}\cdot \delta \vec{\bm{p}}}{\left( 1-\vec{\bm{v}%
}\cdot \hat{\bm{n}}\right) }\right) \right] _{E}^{R},
\label{eq:fre5}
\end{equation}%
where $[z]_{E}^{R}$ means $z|_{R}-z|_{E}$ for an arbitrary
function $z$. If we approximate the logarithm on the right hand
side of (\ref{eq:fre5}), we may write
\begin{equation}
\ln \left( \frac{\nu _{R}}{\nu _{E}}\right) _{gr}=\left[ \left( \delta p^{0}-%
\vec{\bm{v}}\cdot \delta \vec{\bm{p}}\right) \left( 1+\vec{\bm{v}}\cdot \hat{%
\bm{n}}\right) \right] _{E}^{R}.  \label{eq:fre6}
\end{equation}%

In order to calculate (\ref{eq:fre6}), we must evaluate the
perturbations of the photons' momentum, at the emission and
reception points, which can be done by integrating the geodesic
equation. We reproduce here, without details, the results of
\cite{bertotti92}.

The perturbations in the photons' four-momentum can be evaluated
by solving the geodesic equations:
\begin{equation}
\frac{d \delta p^\mu}{d\lambda} +\Gamma^{\mu}_{\alpha\beta}
\left(p_0^\alpha+\delta p^\alpha \right)\left(p_0^\beta+\delta
p^\beta \right)=0. \label{eq:geod11}
\end{equation}
We assume that the gravitational field is everywhere weak, so that
it can be written in the form $g_{\mu \nu }=\eta _{\mu \nu
}+h_{\mu \nu }$, where $h_{\mu \nu }$ is treated as a perturbation
of the Minkowski background $\eta _{\mu \nu }$. Consequently, on
using a linear approximation and supposing that the perturbations
originate only in a small region of space around the source of
gravitational field (this is the case if the impact parameter $b$
is much smaller than the distances of the emitter and receiver:
$b\ll l_{E},l_{R}$), we can write \cite{bertotti92}:
\begin{eqnarray}
\delta p_{\mu }|_{E} &=&-\frac{l_{R}}{2l}\int_{-\infty }^{\infty
}h_{\alpha \beta ,\mu}(\lambda )p_{0}^{\alpha }p_{0}^{\beta
}d\lambda ,
\label{eq:dpmuE} \\
\delta p_{\mu }|_{R} &=&\frac{l_{E}}{2l}\int_{-\infty }^{\infty
}h_{\alpha \beta ,\mu}(\lambda )p_{0}^{\alpha }p_{0}^{\beta
}d\lambda . \label{eq:dpmuR}
\end{eqnarray}%

The integrals (\ref{eq:dpmuE}),(\ref{eq:dpmuR}) must be evaluated
along the unperturbed photons' path, i.e. along the straight line
connecting the emission point and the reception point.\newline

Eq. (\ref{eq:fre6}), together with eqs. (\ref{eq:dpmuE}) and (\ref{eq:dpmuR}%
) are all we need to evaluate, to the desired order, the
gravitational contribution to the Doppler frequency shift.


\section{Applications}\label{sec:application}

After having determined the perturbations of the photons'
momentum, we may apply eq. (\ref{eq:fre6}) for evaluating the
gravitational frequency shift. In the following, we study the
frequency shift induced by a spherical distribution of mass, then
we take into account a quadrupole contribution to the
gravitational field and study the corresponding frequency shift
and, finally, we study the effect of the rotation of the source of
the gravitational field.


\subsection{The weak field around a spherical distribution of mass}

\label{ssec:wfm}

\begin{figure}[top]
\begin{center}
\includegraphics[width=9cm,height=7cm]{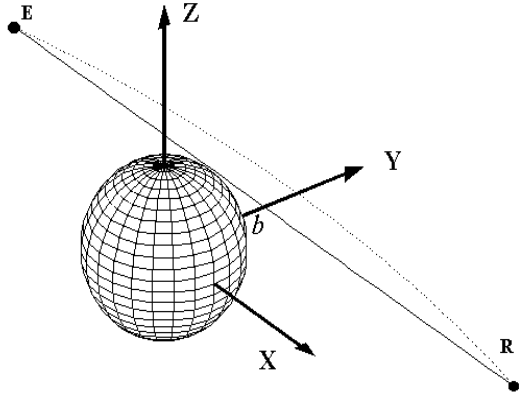}
\end{center}
\caption{{\protect\small Because of the presence of the massive
source, the line linking the emission and reception points $E$,
$R$ bends, and does not follow a straight line.}}
\label{fig:bending}
\end{figure}

Let us consider a spherical mass distribution, and let us suppose
that its gravitational field is weak. The  line element can be
written in the PPN form (up to $O(v^{2})$) \cite{MTW}{\normalsize
\
\begin{equation}
ds^{2}=-\left( 1-\frac{2M}{r}\right) dt^{2}+\left( 1+\frac{2\gamma M}{r}%
\right) \left( dx^{2}+dy^{2}+dz^{2}\right),   \label{eq:wfmass1}
\end{equation}%
} where $r=\sqrt{x^2+y^2+z^2}$, and $\gamma$ is a PPN parameter.
Let us suppose that photons propagate in the plane $z=0$, and that
their unperturbed path is parallel to the $x$ axis, with impact
parameter $b$ \footnote{In fact $b$ is the ratio between two
constants of the motion and coincides with the closest approach
distance from the origin in the case of a flat
space-time.} (see Figure \ref{fig:bending}). In other words, we have%
{\normalsize \
\begin{equation}
\bm{p}_{0}=\left( 1,\hat{\bm{n}}\right) \ \ \ \
\hat{\bm{n}}\parallel \hat{\bm{x}}.  \label{eq:wfmass2}
\end{equation}%
}Considering that we have no change of the time component of the
photons' momentum, since the metric does not depend on time
("conservation of energy"), eq. (\ref{eq:fre6}), up to the lowest
order in $v$ becomes{\normalsize \
\begin{equation}
\ln \left( \frac{\nu _{R}}{\nu _{E}}\right) =\left[ \left( -\vec{\bm{v}}%
\cdot \delta \vec{\bm{p}}\right) \right]
_{E}^{R}=-\vec{\bm{v}}\cdot \delta
\vec{\bm{p}}|_{R}+\vec{\bm{v}}\cdot \delta
\vec{\bm{p}}|_{E}+O(v^{4}). \label{eq:freM1}
\end{equation}%
}Then, according to eqs. (\ref{eq:dpmuE}) and
(\ref{eq:dpmuR}){\normalsize \
\begin{equation}
\delta p^{x}|_{E}=\delta p^{x}|_{R}=0,\ \ \ \delta
p^{z}|_{E}=\delta p^{z}|_{R}=0,  \label{eq:wfmass4}
\end{equation}%
}and the only non null perturbations in the photons' world-line are given by%
\begin{eqnarray}
\delta p^{y}|_{E} &=&-\frac{l_{R}}{2l}\int_{-\infty }^{\infty
}2{h}_{xx,}^{\ \ \
\ y}dx,  \label{eq:dpyME1} \\
\delta p^{y}|_{R} &=&\frac{l_{E}}{2l}\int_{-\infty }^{\infty
}2{h}_{xx,}^{\ \ \ \ y}dx.  \label{eq:dpyMR1}
\end{eqnarray}%
On integrating (\ref{eq:dpyME1}) and (\ref{eq:dpyMR1}) we obtain
{\normalsize \
\begin{equation}
\delta p^{y}|_{E}=\frac{2(1+\gamma )M}{b}\frac{l_{R}}{l},\ \ \
\delta p^{y}|_{R}=-\frac{2(1+\gamma )M}{b}\frac{l_{E}}{l}.
\label{eq:dpyMRE}
\end{equation}%
}

As a consequence, the frequency shift due to a spherical
distribution of mass, turns out to be{\normalsize \
\begin{equation}
\ln \left( \frac{\nu _{R}}{\nu _{E}}\right) _{M}=\frac{2(1+\gamma )M}{b}%
\left[ \frac{(v_{R})_{y}l_{E}+(v_{E})_{y}l_{R}}{l}\right]
+O(v^{4}). \label{eq:freM2}
\end{equation}%
}


\subsubsection{On the physical meaning of the gravitational frequency shift}\label{sssec:mean}

\begin{figure}[top]
\begin{center}
{\normalsize \includegraphics[width=9cm,height=7cm]{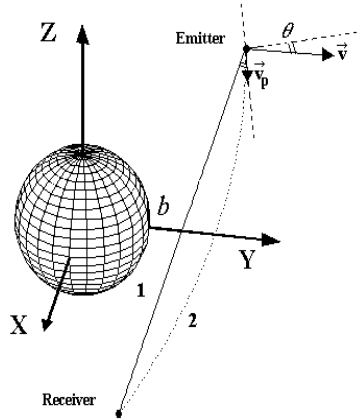} }
\end{center}
\caption{{\protect\small The combined effect of the emitter's
motion and of the gravitational bending of light rays produces a
gravitational Doppler effect.}} \label{fig:velocity}
\end{figure}

What we have seen so far, allows us to give a simple
interpretation of the gravitationally induced frequency
shift.\newline Let us consider an emitter moving with velocity
$\vec{\bm{v}}$ nearby a massive source (see Figure
\ref{fig:velocity}). Let us suppose that the emitter's velocity is
orthogonal to the line connecting it to the receiver. If we forget
the gravitational field, we know that, in this case, there is
no longitudinal Doppler effect (or first order Doppler effect, since it is $%
O(v)$), but there is only a transverse Doppler effect (which is
$O(v^{2})$, i.e. of second order). This depends on the fact that
the velocity of the emitter has no components along the line of
sight. However, we know that one of the effects of the
gravitational field is the bending of the light rays: this means
that, roughly speaking, the light ray propagates along the curve
$2$ and not along the straight line $1$. As a consequence, there
is a component of the emitter's velocity along the actual
direction of propagation of the signal, and its value can easily
be calculated. In fact, the deflection angle $\theta $ in GR (see,
for instance, \cite{straumann}) is given by{\normalsize \
\begin{equation}
\theta =\frac{2M}{b}.  \label{eq:thetaM}
\end{equation}%
}From Figure \ref{fig:velocity}, it is easy to recognize that the
component of the emitter's velocity along the propagation line is
approximately given by{\normalsize \
\begin{equation}
v_{p}\simeq v 2 \theta =v\frac{4M}{b}.  \label{eq:vp1}
\end{equation}%
}To this longitudinal velocity it corresponds a first order Doppler effect%

\begin{equation}
\left\vert \frac{\delta \nu }{\nu _{E}}\right\vert =v_{p},
\label{eq:vp2}
\end{equation}%
or, reintroducing physical units,
\begin{equation}
\left\vert \frac{\delta \nu }{\nu _{E}}\right\vert
=\frac{v}{c}\frac{4GM}{c^2b}, \label{eq:vp22}
\end{equation}
which is in agreement with (\ref{eq:freM2}) if we set $\gamma =1$,
as in GR, and let $l_{E}\ll l_{R}$ (see below). So, this frequency
shift may be explained in terms of the bending of the light rays
due to the gravitational field, and it is a third order
$(O(v^{3}))$ Doppler effect.


\subsection{Quadrupole contribution in weak field approximation}

\label{ssec:wfq}

If the mass distribution is not perfectly spherical (still
preserving an axial symmetry), the quadrupole contribution comes
into play. In this case, the gravitational potential is written in
the form \cite{bertottifarinella90} {\normalsize
\begin{equation}
\phi (\bm{x})=-\frac{M}{r}\left[ 1-J_{2}\left( \frac{R}{r}\right) ^{2}\frac{%
3\cos ^{2}\theta -1}{2}\right] ,  \label{eq:wfq1}
\end{equation}%
}where {\normalsize $R$} is the average radius of the mass
distribution, and $J_{2}=-\frac{Q_{zz}}{MR^{2}}$, contains the
components of the quadrupole tensor $Q_{ij}$.

In particular, the quadrupole contribution to the gravitational
potential is {\normalsize
\begin{equation}
\phi (\bm{x})_{Q}\doteq \frac{M}{r}J_{2}\left( \frac{R}{r}\right) ^{2}\frac{%
3\cos ^{2}\theta -1}{2}.  \label{eq:wfq2}
\end{equation}%
}

Let us work out the quadrupole contribution to the gravitational
Doppler effect. The line element is written in the
form{\normalsize \
\begin{equation}
ds^{2}=-\left( 1+2\phi \right)dt^2 +\left( 1-2\gamma \phi \right)
\left( dx^{2}+dy^{2}+dz^{2}\right) .  \label{eq:wfq3}
\end{equation}%
}As before, we consider photons propagating in the equatorial
plane ($z=0$ or $\theta =\pi /2$) of the source, parallel to the
$x$ axis. This amounts to writing the metric perturbation $h_{xx}$
in the form
\begin{equation}
h_{xx}=\frac{2\gamma M}{r}+J_{2}\frac{\gamma M}{r}\left(
\frac{R}{r}\right) ^{2}=\left( h_{xx}\right) _{M}+\left(
h_{xx}\right) _{Q}. \label{eq:wfq4}
\end{equation}
The  first term, giving the monopole contribution to the frequency
shift, has been calculated before; now we focus on the second
term, expressing the quadrupole effect.

Proceeding as before, we obtain that the only non null perturbations are given by%

\begin{equation}
\delta p^{y}|_{E} = \frac{2J_{2}(1+\gamma )M}{b}\left( \frac{R}{b}\right) ^{2}%
\frac{l_{R}}{l}, \ \ \  \delta p^{y}|_{R}= -\frac{2J_{2}(1+\gamma
)M}{b}\left( \frac{R}{b}\right) ^{2}\frac{l_{E}}{l}.
\label{eq:wfqdpyMRE}
\end{equation}%
Then, from
\begin{equation}
\ln \left( \frac{\nu _{R}}{\nu _{E}}\right) =\left[ \left( -\vec{\bm{v}}%
\cdot \delta \vec{\bm{p}}\right) \right]
_{E}^{R}=-\vec{\bm{v}}\cdot \delta
\vec{\bm{p}}|_{R}+\vec{\bm{v}}\cdot \delta
\vec{\bm{p}}|_{E}+O(v^{4}), \label{eq:wfqfreM1}
\end{equation}%
taking into account the perturbations (\ref{eq:wfqdpyMRE}), we
obtain for the quadrupole contribution{\normalsize \
\begin{equation}
\ln \left( \frac{\nu _{R}}{\nu _{E}}\right) _{Q}=\frac{2J_{2}(1+\gamma )M}{b}%
\left( \frac{R}{b}\right) ^{2}\left[ \frac{(v_{R})_{y}l_{E}+(v_{E})_{y}l_{R}%
}{l}\right]   +O(v^{4}).  \label{eq:wfq6}
\end{equation}%
}


\subsection{The weak field around a rotating distribution of mass}\label{ssec:wfj}

In the case of a rotating source and weak field approximation the
line element, to the lowest meaningful order in PPN form, is
\cite{bertotti92,will05}{\normalsize \
\begin{equation}
ds^{2}=-\left( 1-\frac{2M}{r}\right) dt^{2}+\left( 1+\frac{2\gamma M}{r}%
\right) \left( dx^{2}+dy^{2}+dz^{2}\right)  +\frac{\left( 4\gamma
+4+\alpha _{1}\right) }{2r^{3}}\left( \vec{\bm{x}}\times
\vec{\bm{S}}\right) _{i}dx^{i}dt, \label{eq:wfmj}
\end{equation}%
}where $\vec{\bm{S}}$ is the angular momentum and $\alpha_1$ is
another PPN parameter. As it is well known, the angular momentum
contribution to the gravitational field, in this weak field
context,  is usually referred to as \textit{gravito-magnetic
field} \cite{ruggiero02}.

Since the angular momentum contribution is of the order
$O(v^{3})$, we do expect that the gravito-magnetic contribution
to the gravitational doppler shift is $O(v^{4})$. In particular,
it is possible to show \cite{bertotti92} that this contribution is
expressed by
\begin{equation}
\ln \left( \frac{\nu _{R}}{\nu _{E}}\right) _{GM}=\left[ \left( -\vec{\bm{v}}%
\cdot ^{(3)}\delta \vec{\bm{p}}\right) \right] _{E}^{R}.
\label{eq:fregm1}
\end{equation}%
where $^{(3)}\delta \vec{\bm{p}}$ means that  we have to evaluate
the perturbations up to $O(v^{3})$.

We consider, as before, photons propagating in the equatorial
plane ($z=0$ or $\theta =\pi /2$) of the source, parallel to the
$x$ axis; furthermore, we assume that the angular momentum of the
source is parallel to the $z$ axis: consequently there are no
perturbations of the momentum along the $x$
axis, and the non null perturbations are given by (see \cite{bertotti92})%
{\normalsize \
\begin{eqnarray}
^{(3)}p^{y}|_{E} &=&\frac{l_{R}}{l}\frac{S}{b^{2}}\left( 2(\gamma +1)+\frac{%
\alpha _{1}}{2}\right) ,  \label{eq:dpyjE} \\
^{(3)}p^{y}|_{R} &=&-\frac{l_{E}}{l}\frac{S}{b^{2}}\left( 2(\gamma +1)+\frac{%
\alpha _{1}}{2}\right).  \label{eq:dpyjR}
\end{eqnarray}%
}

So the gravito-magnetic contribution to the frequency shift turns out to be%

\begin{equation}
\ln \left( \frac{\nu _{R}}{\nu _{E}}\right)
_{GM}=\frac{S}{b^{2}}\left(
2(\gamma +1)+\frac{\alpha _{1}}{2}\right) \left[ \frac{%
(v_{R})_{y}l_{E}+(v_{E})_{y}l_{R}}{l}\right]+O(v^{5}) .
\label{eq:frej1}
\end{equation}%

{\normalsize
}

\section{Examples and Numerical Estimates}

{\normalsize \label{sec:estimates}
}

In order to evaluate the magnitude of the frequency shifts studied
above, we consider three cases corresponding to the different
positions of the emitter and receiver with respect to the source
of the gravitational field:

\begin{enumerate}
\item {\normalsize $l_E \ll l_R$ }

\item {\normalsize $l_R \ll l_E$ }

\item {\normalsize $l_E \simeq l_R$ }
\end{enumerate}

The cases above correspond to different actual situations, that
may prove useful for detecting the effect. Namely, case 1 can
occur, say, in the Solar System, when an emitter is on board a
spacecraft orbiting the Sun or one planet, and the receiver is far
away, for instance, on the Earth. Case 2 could correspond to a
source of electromagnetic signals, which is very distant from the
massive source of the gravitational field. For instance, one may
think of a distant star, such as a pulsar, emitting
electromagnetic signals, whose paths pass close to the Sun, and
then are received on the Earth. Finally, case 3 may occur when
both the emitter and the receiver are moving on similar orbits
around a massive body. In the following, we shall evaluate the
magnitude of the frequency shifts for some prototypical
situations.\newline

Before going on, we notice that if we assume that{\normalsize \
\begin{equation}
\nu _{R}=\nu _{E}+\delta \nu ,  \label{eq:ne1}
\end{equation}%
}we may write {\normalsize
\begin{equation}
\ln \left( \frac{\nu _{R}}{\nu _{E}}\right) =\ln \left( \frac{\nu
_{E}+\delta \nu }{\nu _{E}}\right) =\ln \left( 1+\frac{\delta \nu }{\nu _{E}}%
\right) \simeq \frac{\delta \nu }{\nu _{E}},  \label{eq:ne2}
\end{equation}%
}since the expected frequency shift is small.


\subsection{Case 1., $l_{E}\ll l_{R}$}

\label{ssec:orbit}

\begin{table}[top]
\begin{center}
{\normalsize \medskip
\begin{tabular}{l|c|c}
Source & $\ S$ (Kg \ m$^2$ \ s$^{-1}$) & $J_2$ \\
\hline
{\small Sun} & $190.0\times 10^{39}$ & $2\times 10^{-7}$ \\
Earth & $5.86 \times 10^{33}$ & $1.08 \times 10^{-3}$ \\
Jupiter & $4.33 \times 10 ^{38}$ & $1.49 \times 10^{-2}$ \\
Saturn & $6.63 \times 10^{37}$ & $1.62 \times 10^{-2}$ \\
&  &
\end{tabular}
}
\end{center}
\caption{{\protect\small Spin Angular Momenta $S$ and $J_{2}$ for
some of the bodies of the Solar System. The value for the Sun's
spin has been taken from
\protect\cite{Pijpers98}. The ratio of the polar moment of inertia $C$ to $%
MR^{2}$ (2/5=0.4 for a homogeneous sphere) is equal to 0.3308 for
the Earth \protect\cite{earth}; for the giant planets it is around
0.2
\protect\cite{giantpl1, giantpl2}. With respect  to the quadrupole mass moments $%
J_{2}$, the value for the solar oblateness has been taken from
\protect\cite{Pit}. For the Earth see, e.g. \protect\cite{ggm02}.
For the giant planets
see \protect\cite{jac1,jac2}, respectively, and http://ssd.jpl.nasa.gov/sat$%
\_$gravity.html$\#$ref8. }} \label{tab:table}
\end{table}


As we said above, this situation may occur when the emitter is
orbiting a massive source, and the receiver is far away from it.
To fix the ideas, we may think of an emitter on a planet around
the Sun, or on board a spacecraft
orbiting the Sun or some other planets of the Solar System. Provided that%
{\normalsize \
\begin{equation}
l_{E}\ll l_{R},  \label{eq:leminlr1}
\end{equation}%
}we have $l_{R}\simeq l$. As a consequence, the orders of
magnitude of the various contributions that we have studied above
may be evaluated by considering the following expressions, where,
here and henceforth, for the sake of simplicity, the PPN
parameters are set equal to their GR values (i.e. $\gamma=1,
\alpha_1=0$):
\begin{eqnarray}
\left\vert \frac{\delta \nu }{\nu _{E}}\right\vert _{M} &\simeq &\frac{GM}{%
c^{2}b}\frac{v_{E}}{c},  \label{eq:stimaM1} \\
\left\vert \frac{\delta \nu }{\nu _{E}}\right\vert _{Q} &\simeq &\frac{%
J_{2}GM}{c^{2}b}\left( \frac{R}{b}\right) ^{2}\frac{v_{E}}{c},
\label{eq:stimaQ1} \\
\left\vert \frac{\delta \nu }{\nu _{E}}\right\vert _{GM} &\simeq &\frac{GS}{%
c^{3}b^{2}}\frac{v_{E}}{c},  \label{eq:stimaJ1}
\end{eqnarray}%
where $v_{E}$,  considering Keplerian motion, can be approximately
written as{\normalsize \
\begin{equation}
v_{E}\simeq \sqrt{\frac{GM}{R_{0}}},  \label{eq:velkep1}
\end{equation}%
}where $R_{0}$ is the order of magnitude of the semi(major)-axis
of the orbit. Consequently, we may write{\normalsize \
\begin{eqnarray}
\left\vert \frac{\delta \nu }{\nu _{E}}\right\vert _{M} &\simeq &\frac{GM}{%
c^{2}b}\sqrt{\frac{GM}{c^{2}R_{0}}},  \label{eq:stimaM2} \\
\left\vert \frac{\delta \nu }{\nu _{E}}\right\vert _{Q} &\simeq &\frac{%
J_{2}GM}{c^{2}b}\left( \frac{R}{b}\right)
^{2}\sqrt{\frac{GM}{c^{2}R_{0}}},
\label{eq:stimaQ2} \\
\left\vert \frac{\delta \nu }{\nu _{E}}\right\vert _{GM} &\simeq &\frac{GS}{%
c^{3}b^{2}}\sqrt{\frac{GM}{c^{2}R_{0}}}.  \label{eq:stimaJ2}
\end{eqnarray}%
}More explicitly{\normalsize \
\begin{equation}
\left\vert \frac{\delta \nu }{\nu _{E}}\right\vert _{M}\simeq
3,06\times
10^{-9}\left( \frac{M}{M_{\odot }}\right) ^{3/2}\left( \frac{R_{\odot }}{b}%
\right) \left( \frac{R_{\odot }}{R_{0}}\right) ^{1/2},
\label{eq:stimaM3}
\end{equation}%
}

{\normalsize
\begin{equation}
\left| \frac{\delta \nu}{\nu_E} \right|_{Q} \simeq 3,06 \times
10^{-9}
J_2\left(\frac{M}{M_\odot}\right)^{3/2}\left(\frac{R_\odot}{b}%
\right)^{3}\left(\frac{R}{R_\odot}\right)^{2}\left(\frac{R_\odot}{R_0}%
\right)^{1/2},  \label{eq:stimaQ3}
\end{equation}
}

{\normalsize
\begin{equation}
\left| \frac{\delta \nu}{\nu_E} \right|_{GM} \simeq 1,41 \times
10^{-15}
\left(\frac{S}{S_\odot}\right)\left(\frac{R_\odot}{b}\right)^{2}\left(\frac{M%
}{M_\odot}\right)^{1/2}\left(\frac{R_\odot}{R_0}\right)^{1/2}.
\label{eq:stimaJ3}
\end{equation}
}

The orders of magnitude of the frequency shifts (\ref{eq:stimaM3}),(\ref%
{eq:stimaQ3}),(\ref{eq:stimaJ3}) are evaluated in Table
\ref{tab:table1}, where we have considered the Sun and some
planets of the Solar System as
sources of the gravitational field, and we have chosen orbits such that $%
R_{0}\simeq b \simeq R$.

\begin{table}[top]
\begin{center}
{\normalsize \medskip
\begin{tabular}{l|c|c|c}
Source & $\left|\frac{\delta \nu}{\nu_E}\right|_{M}$ &
$\left|\frac{\delta
\nu}{\nu_E}\right|_{Q}$ & $\left|\frac{\delta \nu}{\nu_E}\right|_{GM}$ \\
\hline {\small Sun} & $3.06 \times 10^{-9}$ & $6.12 \times
10^{-16}$ & $1.41 \times
10^{-15} $ \\
Earth & $1.86 \times 10^{-14}$ & $2.02 \times 10^{-17}$ & $3.03
\times
10^{-20}$ \\
Jupiter & $2.28 \times 10 ^{-12}$ & $4.14 \times 10^{-14}$ & $9.50
\times
10^{-17}$ \\
Saturn & $6.28 \times 10^{-13}$ & $1.02 \times 10^{-14}$ & $1.32
\times
10^{-17}$ \\
&  &  &
\end{tabular}
}
\end{center}
\caption{{\protect\small Evaluation of the frequency shifts
induced by the gravitational field for emitters orbiting the
bodies of the Solar System.}} \label{tab:table1}
\end{table}


\subsection{Case 2., $l_{R}\ll l_{E}$}

\label{ssec:far}

Another interesting situation corresponds to a source of
electromagnetic signals, which is indeed very distant from the
massive source of gravitational field. For instance, one may think
of a distant star, such as a pulsar, emitting electromagnetic
signals, whose paths pass close to the Sun, and then are received
on the Earth. In this case, we have
\begin{equation}
l_{R}\ll l_{E},  \label{eq:leminlr11}
\end{equation}%
so that $l_{E}\simeq l$. As a consequence, the velocity which
plays a role is now the one of the receiver, i.e. the Earth-based
observer. To fix the ideas, for a pulsar whose beam passes near
the Sun, $v_{R}$ is nothing but the apparent velocity of the Sun
in the sky, i.e. $v_{R}\simeq 10^{4}\ m/s$. Then{\normalsize \
\begin{eqnarray}
\left\vert \frac{\delta \nu }{\nu _{E}}\right\vert _{M} &\simeq &\frac{GM}{%
c^{2}b}\frac{v_{R}}{c},  \label{eq:stimaM11} \\
\left\vert \frac{\delta \nu }{\nu _{E}}\right\vert _{Q} &\simeq &\frac{%
J_{2}GM}{c^{2}b}\left( \frac{R}{b}\right) ^{2}\frac{v_{R}}{c},
\label{eq:stimaQ11} \\
\left\vert \frac{\delta \nu }{\nu _{E}}\right\vert _{GM} &\simeq &\frac{GS}{%
c^{3}b^{2}}\frac{v_{R}}{c}.  \label{eq:stimaJ11}
\end{eqnarray}%
}On evaluating for $M=M_{\odot },$ $S=S_{\odot }$, we
have{\normalsize \
\begin{eqnarray}
\left\vert \frac{\delta \nu }{\nu _{E}}\right\vert _{M} &\simeq
&7.06\times
10^{-11}\left( \frac{R_{\odot }}{b}\right) ,  \label{eq:stimaM21} \\
\left\vert \frac{\delta \nu }{\nu _{E}}\right\vert _{Q} &\simeq
&7.06\times
10^{-11}J_{2}\left( \frac{R_{\odot }}{b}\right) ^{3},  \label{eq:stimaQ21} \\
\left\vert \frac{\delta \nu }{\nu _{E}}\right\vert _{GM} &\simeq
&3.24\times 10^{-17}\left( \frac{R_{\odot }}{b}\right) ^{2}.
\label{eq:stimaJ21}
\end{eqnarray}%
}It is clear that the most favourable position is near to the
opposition, where the impact parameter is as close as possible to
the radius of the Sun. Near to the opposition, we may
pose{\normalsize \
\begin{equation}
|v_{R}|=\left\vert \frac{db}{dt}\right\vert .  \label{eq:defvrb1}
\end{equation}%
}As a consequence, the estimates of the frequency shifts can now
be written as{\normalsize \
\begin{eqnarray}
\left\vert \frac{\delta \nu }{\nu _{E}}\right\vert _{M} &\simeq &\frac{GM}{%
c^{2}b}\left\vert \frac{db}{cdt}\right\vert ,  \label{eq:stimaM31} \\
\left\vert \frac{\delta \nu }{\nu _{E}}\right\vert _{Q} &\simeq &\frac{%
J_{2}GM}{c^{2}b}\left( \frac{R}{b}\right) ^{2}\left\vert \frac{db}{cdt}%
\right\vert ,  \label{eq:stimaQ31} \\
\left\vert \frac{\delta \nu }{\nu _{E}}\right\vert _{GM} &\simeq &\frac{GS}{%
c^{3}b^{2}}\left\vert \frac{db}{cdt}\right\vert .
\label{eq:stimaJ31}
\end{eqnarray}%
}We point out that the frequency shifts (\ref{eq:stimaM31}) and (\ref%
{eq:stimaJ31}) were already obtained in a previous work
\cite{tartaglia04}, using a different approach based on the
calculation of the gravitational induced time delay
(\textquotedblleft Shapiro time delay\textquotedblright );
furthermore, (\ref{eq:stimaM31}) corresponds to the recent
measurements performed by Bertotti and collaborators
\cite{bertotti03} of the PPN $\gamma $ parameter by means of radio
ranging to the Cassini spacecraft.


\subsection{Case 3., $l_{E}\simeq l_{R}$}

\label{ssec:equal}

This situation may occur, for instance, when both the emitter and
the receiver are orbiting the source of the gravitational field,
and their orbits have similar dimensions. One might think of two
satellites communicating with each other and orbiting the Earth.
To fix the ideas, one could consider two GPS satellites, even
though in their present configuration they can communicate with
the Earth-based stations only. However, the satellites of the
forthcoming GALILEO positioning system, at least in their second
generation, could be able to communicate with one another, so it
is useful to evaluate the magnitude of the effect for this
possible physical situation. In order to evaluate the magnitude of
the effect, we consider two satellites orbiting the Earth, on the
same circular orbit with radius $R_{c}\simeq 26.6\times 10^{6}\
m$; we also assume that the orbit lays in the equatorial plane of
the Earth. Furthermore, we calculate the frequency shift in the
most favourable position of the two satellites, i.e. when $b
\simeq R_{\oplus }$. As a consequence, we may write

\begin{eqnarray}
\left\vert \frac{\delta \nu }{\nu _{E}}\right\vert _{M} &\simeq &\frac{%
GM_{\oplus }}{c^{2}R_{\oplus }}\sqrt{\frac{GM_{\oplus
}}{c^{2}R_{c}}},
\label{eq:stimaM12} \\
\left\vert \frac{\delta \nu }{\nu _{E}}\right\vert _{Q} &\simeq &\frac{%
J_{2}GM_{\oplus }}{c^{2}R_{\oplus }}\sqrt{\frac{GM_{\oplus
}}{c^{2}R_{c}}},
\label{eq:stimaQ12} \\
\left\vert \frac{\delta \nu }{\nu _{E}}\right\vert _{GM} &\simeq &\frac{%
GS_{\oplus }}{c^{3}R_{\oplus }^{2}}\sqrt{\frac{GM_{\oplus
}}{c^{2}R_{c}}}. \label{eq:stimaJ12}
\end{eqnarray}%
Numerically evaluating the formulas, we obtain
\begin{eqnarray}
\left\vert \frac{\delta \nu }{\nu _{E}}\right\vert _{M} &\simeq
&2.80\times
10^{-16},  \label{eq:stimaM22} \\
\left\vert \frac{\delta \nu }{\nu _{E}}\right\vert _{Q} &\simeq
&3.03\times
10^{-19},  \label{eq:stimaQ22} \\
\left\vert \frac{\delta \nu }{\nu _{E}}\right\vert _{GM} &\simeq
&1.44\times 10^{-22}.  \label{eq:stimaJ22}
\end{eqnarray}%

As a general comment, we may say that, in all cases analyzed
above, the gravito-magnetic contribution to the frequency shift is
much lower than the other ones, so it would hardly be detected.
Also, the value of the $J_2$ (Table \ref{tab:table}) for the
bodies of the Solar System, makes the quadrupole contribution much
smaller than the monopole one.


\section{Discussion and Conclusions}

\label{sec:conclusions}

We have shown that the  relative frequency shifts originating from
the gravitational bending of light rays range from $10^{-9}$ to
$10^{-22}$. The measurements of these frequency shifts, could
allow, in principle, to  estimate  the PPN parameters that we have
introduced in our formulas, or to measure the gravitational field
of the celestial bodies.  Actually, as we pointed out before, a
recent estimate of the $\gamma $ parameter, was performed, by
means of the measurement of the frequency shift of radio photons
to and from the Cassini spacecraft as they passed near the Sun
\cite{bertotti03}. We remember that, in GR, $\gamma =1,\alpha_1
=0$, and the  most recent values of the PPN parameters are
reported in \cite{will05}.

Of course, in order to evaluate the reliability of actual
measurements of these effects, a careful and detailed analysis of
the error budget is required. In this work, however, we are just
interested in pointing out the possibility of detection of some of
them, on the bases of the knowledge of the current accuracy in
frequency shift measurements.

We see that, in order to measure the quadrupole contribution, an
accuracy of $10^{-14}$ at least is required, while the
gravito-magnetic contribution requires an accuracy of $10^{-15}$.
Nowadays, it is  known that an atomic standard stable to $1$ part
in $10^{16}$ is  available \cite{ioni}, so that such a fractional
frequency stability is not unreasonable for future experiments
with atomic clocks. Indeed, the giant planets, Jupiter and Saturn,
are, at least in principle, candidates for such measurements: in
fact, the estimates of their quadrupole moment seem to be in the
actual or foreseeable accuracy of the instruments. Furthermore,
they do not have an electromagnetic activity as intense as the one
of the Sun: so, again, at least in principle, it would not bias
the radio communications.

We must remember that in an actual measurement, besides the
gravitationally induced terms, also the ordinary Doppler effect
($O(v)$), the second order Doppler effect ($O(v^{2})$ and the
gravitational redshift ($O(v^{2})$) contribute to the Doppler
signal (in particular, the latter two effects were
tested using a space-borne hydrogen maser carried aboard a rocket \cite{gpa}%
). In order to measure the  frequency shifts induced by the
gravitational bending of light rays, these "ordinary" Doppler
contributions must be subtracted: to this end, a very accurate
knowledge of the orbits of the spacecrafts or planets is required.
On the other hand, the different dependence on the geometric
parameters (such as $b$) of the various contributions, may suggest
a way of discriminating them; in the case of the gravito-magnetic
contribution to the frequency shift, the antisymmetry of the
effect with respect to the rotation axis of the source may help to
disentangle the extremely weak signal from the remaining bigger
shifts due to other causes \cite{tartaglia04}.

In conclusion, we have shown that the Doppler effects induced by
the gravitational bending of light rays, though small, deserve
further analysis. If measured, these effects may allow us to put
limits on the values of the PPN parameters and, on the other hand,
they can be used as probes to understand the nature and explore
the physical properties of the celestial bodies.



\end{document}